\documentstyle[aps, pre, preprint]{revtex}

\begin{document}
\title{Longitudinal wave instability in magnetized high correlation dusty
plasma{\small \footnotemark[2]  }}
\author{XIE Bai-Song$^{1,2,3}$}
\address{{\small $^1$The Key Laboratory of Beam Technology and Materials}\\
{\small Modification of Ministry of Education,}\\
{\small Beijing Normal University, Beijing 100875, China}\\
{\small $^{2}$Institute of Low Energy Nuclear Physics,}\\
{\small Beijing Normal University, Beijing 100875, China}\\
{\small $^3$Beijing Radiation Center, Beijing 100875, China}}
\maketitle
\date{\empty}

\begin{abstract}
Low frequency longitudinal wave instability in magnetized high correlation
dusty plasmas is investigated. The dust charging relaxation is taken into
account. It is found that the instablity of wave is determined significantly
by the frequency of wave, the dust charging relaxation, the shear viscosity
and viscoelastic relaxation time, the coupling parameter of high correlation
of dust as well the strength of magnetic field.

\vskip 2mm PACS number(s): 52.25.Zb, 52.25.Vy, 52.35.Mw

\vskip 2mm
\end{abstract}

\bigskip \footnotetext{$^{\dag }$This work was supported by the National
Natural Science Foundation of China under Grants Nos. 10275007 and
No.10135010, and partially by the Research Fund for the Doctoral Program of
Higher Education (No 20010027005) and the Distinguished Young Teacher Fund
of Ministry of Education of China.}

Low frequency longitudinal volume dust waves have been extensively studied
for weakly coupled and unmagnetized dusty plasmas \cite{rao,bar,mer,ang,xie1}%
. Recently the corresponding research on dust waves have been also extended
to the strongly coupled \cite{ros,kaw,xie2,xie3} as well the magnetized
dusty plasmas \cite{sal,cra1,shu,xie4}. In this letter we study the
instability of longitudinal dust wave in the strongly coupled dusty plasmas.
The stationary external magnetic field is applied. Some important factors
such as the high-correlation on dusts, the wave-number dependence of the
elastic relaxation time and effect of dust charging relaxation are included.
It is shown that there are some frequency regime for the occurence of
longitudinal dust wave instability because of magnetized effect and the
charging collisions between the dusts and the plasma particles. The regime
of wavenumber that longitudinal wave\ is allowed depend also on the
competition between shear viscosity and applied magnetic field strength. The
behaviors of dust wave instability is also modified significantly in
magnetized dusty plasmas by magnetized effect.

\bigskip The applied external magnetic field is assumed along z-direction as 
${\bf B=}B{\bf z}$. Because we consider the longitudinal waves, we just need
one of the Maxwell equations, i.e. Poisson equation, as 
\begin{equation}
\nabla ^{2}\varphi =-4\pi e(\delta n_{i}-\delta n_{e}-Z_{d0}\delta
n_{d}-n_{d0}\delta Z_{d}),  \label{e1}
\end{equation}%
where $\varphi $ is the electrostactic potential, $n_{\alpha 0}$ and $\delta
n_{\alpha }$ are the steady and perturbation density of species $\alpha
=e,i,d$ (electrons, ions and dusts respectively), $Z_{d0}$ and $\delta Z_{d}$
are the steady and perturbation charge numbers of dust negative charge as $%
q_{d}=-eZ_{d0}-e\delta Z_{d}$. Furthermore, the overall charge neutrality
condition holds in the steady state, $n_{i0}=n_{e0}+Z_{d0}n_{d0}$. For
longitudinal modes, one has to assumes that the wave propagates nearly
perpendicular to the magnetic field. For simplicity we shall consider wave
along $x$. The perturbations associated with the problem are assumed as in
the form of $\sim \exp [i(kx-\omega t)]$. Then the perturbed electrons
(ions) number density in magnetized plasmas can get by ordinary fluid
equation. On the other hand in terms of a generalized viscoelastic
hydrodynamic model \cite{ich} the perturbed density for the dusts in
magnetized plasmas are also obtained. In addition to the dust charging
equation by orbit-motion-limited (OML)\ theory the problem would be closed
by Eq.(\ref{e1}). In Ref.\cite{xie4} we have obtained the general dispersion
relation for low frequency longitudinal dust waves in strongly coupled
magnetized dusty plasmas (please refer to Eq.(11) of Ref.\cite{xie4}) 
\begin{equation}
{\cal A}\frac{\omega _{pi}^{2}}{k^{2}V_{Ti}^{2}+\Omega _{i}^{2}}+{\cal B}%
\frac{\omega _{pe}^{2}}{k^{2}V_{Te}^{2}+\Omega _{e}^{2}}+\epsilon _{pd}=0
\label{e2}
\end{equation}%
where where $\omega _{pe,pi}=\sqrt{4\pi n_{e0,i0}e^{2}/m_{e,i}}$, $n_{e0,i0}$%
, $m_{e,i}$, $V_{Te,Ti}=\sqrt{T_{e,i}/m_{e,i}}$ and $\Omega
_{e,i}=eB/m_{e,i}c$ are the chracteristic oscillation frequencies,
unperturbed number densities, masses, thermal velocities and the
gyro-frequencies of the electrons and ions, respectively, 
\begin{equation}
{\cal A}=1+\frac{P}{P+1}\frac{(\tau +z)}{z(1+\tau +z)}  \label{e3}
\end{equation}%
and 
\begin{equation}
{\cal B}=1+\frac{P(\tau +z)}{z(1+\tau +z),}  \label{e4}
\end{equation}%
are two parameters which are associated with the dust charging, and 
\begin{equation}
\epsilon _{pd}=1-\frac{\omega _{pd}^{2}}{\omega \lbrack \omega +{\rm i}\eta
_{d1}(k,\omega )]-\gamma _{d}\mu _{d}k^{2}V_{Td}^{2}-\Omega _{d}^{2}/[1+\eta
_{d2}(k,\omega )]}  \label{e5}
\end{equation}%
is the dielectric function of the dust fluid. The other quantities are $%
P=Z_{d0}n_{d0}/n_{e0}$, $\tau =T_{i}/T_{e}$ is the ratio of ion temperature
to the electron temperature, and $z=Z_{d0}e^{2}/aT_{e}$ is the normalized
dust-dust interaction potential energy, $\omega _{pd}=\sqrt{4\pi
n_{d0}Z_{d0}^{2}e^{2}/m_{d}}$, $n_{d0}$, $m_{d}$, $V_{Td}=\sqrt{T_{d}/m_{d}}$
and $\Omega _{d}=Z_{d0}eB/m_{d}c$ are the chracteristic oscillation
frequency, unperturbed number density, mass, thermal velocity and the
gyro-frequency of the dust, especially two quantities%
\begin{equation}
\eta _{d1}(k,\omega )=\frac{(\zeta +4\eta /3)k^{2}}{m_{d}n_{d0}(1-{\rm i}%
\omega \tau _{m})},  \label{7a}
\end{equation}%
and 
\begin{equation}
\eta _{d2}(k,\omega )=\frac{\eta k^{2}}{m_{d}n_{d0}(1-{\rm i}\omega \tau
_{m})},  \label{e7b}
\end{equation}%
are associated with the bulk viscosity $\zeta $, the shear viscosity $\eta $
and the viscoelstic relaxation time $\tau _{m}$ for highly correlated dusts.
Note that $\gamma _{d}$ is the adiabatic index and $\mu _{d}=1+u(\Gamma
)/3+(\Gamma /9)\partial _{\Gamma }u(\Gamma )$ is the compressibility and, $%
u(\Gamma )$, the normalized correlation energy, or the excess internal
energy, of the system \cite{ich}. For example when Coulomb coupling
parameter $\Gamma =(Z_{d}e)^{2}/a_{d}T_{d}$, ($-Z_{d}e$ is the dust charge, $%
a_{d}$ the interdust distance and $T_{d}$ the dust temperature) becomes high
enough, the negative dispersion occurs for dust waves \cite{xie3}.. For
weakly coupled dusts ($\Gamma <1$), or under certain conditions also for
strongly coupled dusts we have $u(\Gamma )\approx -\sqrt{3}\Gamma ^{3/2}/2$ %
\cite{kaw,xie2,ich}. In practice $u(\Gamma )$ is usually obtained by fitting
results from experiments and/or Monte Carlo or Molecular Dynamics
simulations. Accordingly, fitting the internal-energy results from a liquid
simulation one obtains $u(\Gamma )\approx -0.90\Gamma $ for$1\leq \Gamma
\leq 200$. Other quantities such as the transport coefficients $\zeta
(\Gamma )$, $\eta (\Gamma )$, and $\tau _{m}(\Gamma )$ are usually too
complex to be expressed analytically, so that often simplifying models are
employed \cite{ich}.

Introducing the dimensionless quantities $\omega =\omega /\omega _{pd}$, $%
\Omega _{d}=\Omega _{d}/\omega _{pd}$, $k=k\lambda _{p}$ (where $\lambda
_{p}=(\lambda _{i}^{-2}+\lambda _{e}^{-2})^{-1/2}$ the plasma Debye scale
length, $\lambda _{i}=\sqrt{T_{i}/4\pi n_{i0}e^{2}}$ is the ion Debye length,%
$\lambda _{e}=\sqrt{T_{e}/4\pi n_{e0}e^{2}}$ is the electron Debye length), $%
\tau _{m}=\tau _{m}\omega _{pd}$, $\eta _{1}^{\ast }={(\zeta +4\eta /3)}/{%
m_{d}n_{d0}\omega _{pd}a_{d}^{2}}$, $\eta _{2}^{\ast }={\eta }/{%
m_{d}n_{d0}\omega _{pd}a_{d}^{2}}$, and $\lambda _{e,i,d,p}=\lambda
_{e,i,d,p}/\lambda _{p}$, we can express the general dispersion relation of
dust waves Eq. (\ref{e2}) as 
\[
\omega ^{2}-\gamma _{d}\mu _{d}k^{2}\lambda _{d}^{2}+\frac{{\rm i}\omega
k^{2}\kappa ^{2}\eta _{1}^{\ast }}{1-{\rm i}\omega \tau _{m}}-\frac{\Omega
_{d}^{2}}{1+{\rm i}k^{2}\kappa ^{2}\eta _{2}^{\ast }/(1-{\rm i}\omega \tau
_{m})\omega } 
\]%
\begin{equation}
=\frac{1}{1+{\cal C}(k)}.  \label{e15}
\end{equation}%
where $\kappa \equiv a_{d}/\lambda _{p}$ is the ratio of interdust distance
to plasma Debye length and 
\begin{equation}
{\cal C}(k)={\cal A}\frac{1}{k^{2}\lambda _{i}^{2}+{\cal E}_{i}}+{\cal B}%
\frac{1}{k^{2}\lambda _{e}^{2}+{\cal E}_{e}}.  \label{e16}
\end{equation}%
with 
\[
{\cal E}_{i}=\frac{\Omega _{i}^{2}}{\omega _{pi}^{2}},{\cal E}_{e}=\frac{%
\Omega _{e}^{2}}{\omega _{pe}^{2}}. 
\]

For weak magnetized plasmas or unmagnetized plasmas we have ${\cal C}%
(k)\approx {\cal A}/k^{2}$ and $\Omega _{d}\ll 1$ so that Eq.(\ref{e15})
will be reduced to the former extensively studied cases \cite{kaw,xie3}.
However for relatively strong magnetized plasmas we have 
\begin{equation}
{\cal C}(k)\approx \frac{{\cal A}}{k^{2}+\Omega _{i}^{2}/\omega _{pi}^{2}}
\label{e18}
\end{equation}%
here we ignore the contribution from electrons since usually $\lambda
_{e}\gg \lambda _{i}$ and .$\Omega _{e}\ll \omega _{pe}$. In the following
we would study two typical cases, i.e., the kinetic regime of $\omega \tau
_{m}\ll 1$ and hydrodynamic regime of $\omega \tau _{m}\gg 1$.

In the kinetic regime of $\omega \tau _{m}\ll 1$ from Eq.(\ref{e15}) we have 
\begin{equation}
\omega ^{2}+{\rm i}\omega k^{2}\kappa ^{2}\eta _{1}^{\ast }-\frac{\Omega
_{d}^{2}}{1+{\rm i}k^{2}\kappa ^{2}\eta _{2}^{\ast }/\omega }-\gamma _{d}\mu
_{d}k^{2}\lambda _{d}^{2}-\frac{1}{1+{\cal C}(k)}=0.  \label{e19}
\end{equation}%
Furthermore we discuss two limiting cases.

(a) For long-wavelength wave, i.e. the wavelength is much longer than the
inter-dust distance, $k\kappa \ll 1$, we have 
\begin{equation}
\omega ^{2}+{\rm i}\omega k^{2}\kappa ^{2}\left( \eta _{1}^{\ast }+\frac{%
\Omega _{d}^{2}}{\omega ^{2}}\eta _{2}^{\ast }\right) -\Omega
_{d}^{2}-\gamma _{d}\mu _{d}k^{2}\lambda _{d}^{2}-\frac{{\cal A}}{%
(V_{Ai}^{2}+{\cal A})^{2}}k^{2}-\frac{V_{Ai}^{2}}{V_{Ai}^{2}+{\cal A}}=0.
\label{e20}
\end{equation}%
where $V_{Ai}=\Omega _{i}/\omega _{pi}$ is the normalized Alfven velocity by
light speed $c$. From this we can always get $k=k_{r}+{\rm i}k_{i}$ that
means there exist wave instablity. Now we look when we can get the pure
exponent-type unstable wave which needs $k_{r}=0$. In fact the frequency of
wave is in the same order of or less much than the dusty plasma frequency $%
\omega _{pd}$ so that the normalized $\omega \leq O(1)$ or $\ll 1$. On the
other hand $\Omega _{d}\leq O(\omega )$ or $\ll \omega $ and $\eta
_{2}^{\ast }\sim O(\eta _{1}^{\ast })$ or $<\eta _{1}^{\ast }$. Thus if $%
\eta _{1}^{\ast }\sim O(1)$ or $<1$ then the second term of left hand in Eq.(%
\ref{e20}) can be ignored. For this case where the shear effect is not
strong we get 
\begin{equation}
k^{2}=\frac{\omega ^{2}-\Omega _{d}^{2}-V_{Ai}^{2}/(V_{Ai}^{2}+{\cal A})}{%
{\cal A}/(V_{Ai}^{2}+{\cal A})^{2}+\kappa ^{2}(1-0.4\Gamma )/3\Gamma }.
\label{e21}
\end{equation}%
Usually $\Omega _{i}\ll \omega _{pi}$ and 
\begin{equation}
\omega >\omega _{H}\approx \sqrt{\Omega _{d}^{2}+\frac{\Omega
_{i}^{2}/\omega _{pi}^{2}}{{\cal A}}}  \label{e22}
\end{equation}%
that is a hybrid waves of dust and ions in the value of less than the dust
plasma frequency. For $\tau \ll 1$, $z\sim O(1)$, in general ${\cal A}%
\approx 1$ to $2$ then it leads to that when $\kappa >\kappa _{c}=\sqrt{15/2%
{\cal A}}$ and $\Gamma >\Gamma _{c}=2.5\kappa ^{2}/(\kappa ^{2}-\kappa
_{c}^{2})$ then $k^{2}<0$ that indicates the existence of pure imaginary
part of wavenumber. For example for ${\cal A}=2,$ $\kappa _{c}\approx 1.9365$%
, then for $\kappa =2,3,4,5$ we have correspondingly $\Gamma _{c}\approx
40,4.3,3.3$ and $3$ respectively.

Now let us disscuses the condition of $\eta _{1}^{\ast }\leq O(1)$ and $%
\omega \tau _{m}\ll 1$ for the limitation of $\Gamma $. In general it is
difficult to obtain exact expressions for $\tau _{m}$, $\eta _{1}$ and $\eta
_{2}$ etc. However, approximate models can be constructed from the results
of theories and numerical simulations. For example, for the range $10\leq
\Gamma \leq 200$, one can obtain from Table 5 and Fig.~33 of Ref. \cite{ich}
the dependence $\eta _{1}^{\ast }\approx 0.02\sqrt{\Gamma }$ from a one
component plasma theory. On the other hand in terms of $\eta _{1}^{\ast
}/\tau _{m}=\kappa ^{-2}\lambda _{d}^{2}\left( 1-\gamma _{d}\mu
_{d}+4u(\Gamma )/15\right) $ we have $\tau _{m}\approx 3\sqrt{\Gamma }/8$.
Obviously in this range $\eta _{1}^{\ast }\leq O(1)$ and $\omega \tau
_{m}\ll 1$ are both hold for $\omega _{H}<\omega \ll 1$.

Now we have concluded that in this case, long-wavelength wave in the kinetic
regime of $\omega \tau _{m}\ll 1$, the spatial instability occurs for
moderate and high strong-coupling, weak viscosity and moderate
viscoelasticity of dust, hybrid order of low-frequency of wave and moderate
inter-dust distance.

(b) For short-wavelength wave, i.e. the wavelength is in the same order of
or less than the inter-dust distance, $k\kappa \leq O(1)$, and $\eta
_{2}^{\ast }\leq O(\eta _{1}^{\ast })\ll 1$, we have 
\begin{equation}
\omega ^{2}-\gamma _{d}\mu _{d}k^{2}\lambda _{d}^{2}-1=0  \label{e23}
\end{equation}%
and therefore 
\begin{equation}
k^{2}=\frac{\omega ^{2}-1}{\kappa ^{2}(1-0.4\Gamma )/3\Gamma }.  \label{e24}
\end{equation}%
That means the wave instablity can occur only if $\Gamma <2.5$ for $\omega
<1 $ and $10>\Gamma >2.5$ for $\omega >1$ and $O(1)$. In later case $\Gamma
<10$ is need for meet the conditions of weak shear viscosity $\eta
_{1}^{\ast }\approx 0.02\sqrt{\Gamma }\ll 1$ as well $\tau _{m}\approx 3%
\sqrt{\Gamma }/8\ll O(1).$The wave instability occurs independently to the
dust changing relaxation and the applied magnetic strength. Certainly $%
\Omega _{d}/\omega _{pd}\ll 1$ should not be breaked.

Now we have concluded that in this case, short-wavelength wave in the
kinetic regime of $\omega \tau _{m}\ll 1$, the spatial instability occurs
for moderate strong-coupling, very weak viscosity, weak viscoelasticity of
dust, dusty-plasma characteristic oscillation order of frequency of wave and
has nothing to do with inter-dust distance.

In the hydrodynamic regime of $\omega \tau _{m}\gg 1$ we have 
\begin{equation}
\omega ^{2}-k^{2}\kappa ^{2}\frac{\eta _{1}^{\ast }}{\tau _{m}}-\frac{\Omega
_{d}^{2}}{1-k^{2}\kappa ^{2}(\eta _{2}^{\ast }/\tau _{m}\omega ^{2})}-\gamma
_{d}\mu _{d}k^{2}\lambda _{d}^{2}-\frac{1}{1+{\cal C}(k)}=0.  \label{e25}
\end{equation}%
Similarly we discuss two limiting cases.

(a) For long-wavelength wave, i.e. the wavelength is much longer than the
inter-dust distance, $k\kappa \ll 1$, Eq.(\ref{e25}) would be reduced as 
\begin{equation}
\omega ^{2}-k^{2}\kappa ^{2}\frac{\eta _{1}^{\ast }}{\tau _{m}}-\Omega
_{d}^{2}\left( 1+\frac{\eta _{2}^{\ast }}{\tau _{m}}\frac{k^{2}\kappa ^{2}}{%
\omega ^{2}}\right) -\gamma _{d}\mu _{d}k^{2}\lambda _{d}^{2}-\frac{{\cal A}%
}{(V_{Ai}^{2}+{\cal A})^{2}}k^{2}-\frac{V_{Ai}^{2}}{V_{Ai}^{2}+{\cal A}}=0.
\label{e26}
\end{equation}%
We know that In the long-wavelength limit the $k$ dependence of $\tau _{m}$
may be modeled by \cite{ich} $\eta _{1}^{\ast }/\tau _{m}=\kappa
^{-2}\lambda _{d}^{2}\left( 1-\gamma _{d}\mu _{d}+4u(\Gamma )/15\right) $.
On the other hand $(\Omega _{d}^{2}/\omega ^{2})(\eta _{2}^{\ast }/\tau
_{m})\sim O(\eta _{1}^{\ast }/\tau _{m})$. Therefore one can get the
wavenumber of dust wave as 
\begin{equation}
k^{2}\approx \frac{\omega ^{2}-\Omega _{d}^{2}-V_{Ai}^{2}/(V_{Ai}^{2}+{\cal A%
})}{{\cal A}/(V_{Ai}^{2}+{\cal A})^{2}+\kappa ^{2}(1-0.3\Gamma )/\Gamma }
\label{e27}
\end{equation}%
which depends on magnetic field and other parameters. Similarly in the case $%
\Omega _{i}\ll \omega _{pi}$ and 
\begin{equation}
\omega >\omega _{H}\approx \sqrt{\Omega _{d}^{2}+\frac{\Omega
_{i}^{2}/\omega _{pi}^{2}}{{\cal A}}}  \label{e28}
\end{equation}%
we have, therefore, when $\kappa >\kappa _{c}\approx \sqrt{10/3{\cal A}}$
and $\Gamma >\Gamma _{c}\approx 3.3\kappa ^{2}/(\kappa ^{2}-\kappa
_{c}^{2}), $ then $k^{2}<0$. For example for ${\cal A}=2,$ $\kappa
_{c}\approx 1.3$, then for $\kappa =1.4,2,3,4,5$ we have correspondingly $%
\Gamma _{c}\approx 22.0,5.7,4.0,3.7$ and $3.5$ respectively.

It is worthy to note what does mean $\omega \tau _{m}\gg 1$. In general $%
\omega \ll 1$ or $\sim O(1)$ so that $\tau _{m}\gg 1$ is at least required.
That means the highly strong-coupling is needed.

Now we have concluded that in this case, long-wavelength wave in the
hydrodynamic regime of $\omega \tau _{m}\gg 1$, the spatial instability
occurs for highly strong-coupling, highly strong viscoelasticity of dust,
hybrid order or/and dust-oscillation order frequency of wave and moderate
inter-dust distance.

(b) For short-wavelength wave, i.e. the wavelength is in the same order of
or less than the inter-dust distance, $k\kappa \leq O(1)$, and, when $\Omega
_{d}\ll \sqrt{\eta _{2}^{\ast }/\tau _{m}}$ for moderate or/and strong shear
dusty plasmas, we have 
\begin{equation}
\omega ^{2}-k^{2}\kappa ^{2}\frac{\eta _{1}^{\ast }}{\tau _{m}}-\gamma
_{d}\mu _{d}k^{2}\lambda _{d}^{2}-1=0  \label{29}
\end{equation}%
or equivalently 
\begin{equation}
k^{2}=\frac{\omega ^{2}-1}{\kappa ^{2}(1-0.24\Gamma )/3\Gamma }  \label{30}
\end{equation}%
that shows the wavenumber $k^{2}<0$ for $\Gamma <4.16$ when $\omega <1$ and
otherwise $\Gamma >4.16$ when $\omega >1$. Obviously the former case is not
realized for the requirement of $\tau _{m}\gg 1$. Certainly the later case
holds for very highly strong coupling dusty system. On the other hand, when $%
\Omega _{d}\gg \sqrt{\eta _{2}^{\ast }/\tau _{m}}$ for very weak shear dusty
plasma, we have not get the existence of pure unstable wave because of its
contradictive to the requirement of $\omega \tau _{m}\gg 1$.

Now we have concluded that in this case, short-wavelength wave in the
hydrodynamic regime of $\omega \tau _{m}\gg 1$, the spatial instability
occurs for highly strong-coupling, moderate or/and strong viscosity, highly
strong viscoelasticity of dust, dusty-plasma oscillation order frequency of
wave and has nothing to do with inter-dust distance.

Finally let us discusse briefly that the condition for the occurence of wave
instability in the regime of $\omega \tau _{m}\sim O(1)$. In this case the
occurence condition of wave instability is almost same as in the kinetic
regime of $\omega \tau _{m}\ll 1$ except $\tau _{m}\sim O(1)$ add limitation
to $\Gamma $. In fact in the middle regime of $\omega \tau _{m}\sim O(1)$ we
have 
\begin{equation}
\omega ^{2}-k^{2}\kappa ^{2}\frac{\eta _{1}^{\ast }}{\tau _{m}}\frac{1-{\rm i%
}}{2}-\frac{\Omega _{d}^{2}}{1-k^{2}\kappa ^{2}(\eta _{2}^{\ast }/\tau
_{m}\omega ^{2})(1-{\rm i})/2}-\gamma _{d}\mu _{d}k^{2}\lambda _{d}^{2}-%
\frac{1}{1+{\cal C}(k)}=0.  \label{32}
\end{equation}%
For very small $k\kappa \ll 1$, we have 
\begin{equation}
\omega ^{2}-\frac{1-{\rm i}}{2}k^{2}\kappa ^{2}\left( \frac{\eta _{1}^{\ast }%
}{\tau _{m}}+\frac{\Omega _{d}^{2}}{\omega ^{2}}\frac{\eta _{2}^{\ast }}{%
\tau _{m}}\right) -\Omega _{d}^{2}-\gamma _{d}\mu _{d}k^{2}\lambda _{d}^{2}-%
\frac{{\cal A}}{(V_{Ai}^{2}+{\cal A})^{2}}k^{2}-\frac{V_{Ai}^{2}}{V_{Ai}^{2}+%
{\cal A}}=0.  \label{33}
\end{equation}%
since $k\kappa \ll 1$ and $\eta _{1}^{\ast }/\tau _{m}\approx 0.05$ so that
the second term is ignored in Eq.(\ref{33}) and it is recover to the Eq.(\ref%
{e21}). For large $k\kappa \sim O(1)$, we have 
\begin{equation}
\omega ^{2}-\frac{1-{\rm i}}{2}k^{2}\kappa ^{2}\frac{\eta _{1}^{\ast }}{\tau
_{m}}+(1+{\rm i})\frac{\Omega _{d}^{2}\omega ^{2}}{k^{2}\kappa ^{2}(\eta
_{2}^{\ast }/\tau _{m})}-\gamma _{d}\mu _{d}k^{2}\lambda _{d}^{2}-1=0.
\label{34}
\end{equation}%
When $\eta _{1}^{\ast }/\tau _{m}\ll 1$ and $\Omega _{d}\ll \sqrt{\eta
_{2}^{\ast }/\tau _{m}}$, it leads to 
\begin{equation}
\omega ^{2}-\gamma _{d}\mu _{d}k^{2}\lambda _{d}^{2}-1=0  \label{35}
\end{equation}%
which is same as Eq.(\ref{e23}). However $\omega \tau _{m}\sim O(1)$
indicates that the occurence condition of wave instability in this case
requires that the wave frequency is moderately increased and the
strong-coupling is moderately decreased, compared to the case of $\omega
\tau _{m}\ll 1$.

In summary, we have investigated the low frequency longitudinal wave
instability in strongly coupled dusty plasmas in presence of magnetized
field. The effect of dust charging relaxation is taken into accoun. It is
found that the occurence of instability of wave is determined significantly
by the frequency of perturbation wave, the dust charging relaxation, the
shear viscosity and viscoelastic relaxation time, the coupling parameter of
high-correlated dusts as well the strength of magnetic field.


\begin{references}
\bibitem{rao} N. N. Rao, P. K. Shukla, and M. Y. Yu, Planet. Space Sci. {\bf %
38}, 543 (1990).

\bibitem{bar} A. Barkan, R. L.Merlino, and N. D'Angelo, Phys. Plasmas {\bf 2,%
} 3563 (1995).

\bibitem{mer} R. L. Merlino, A. Barkan, C. Thompson, and N. D'Angelo, Phys.
Plasmas {\bf 5,} 1607 (1998).

\bibitem{ang} N. D'Angelo, Phys. Plasmas {\bf 5}, 3155 (1998).

\bibitem{xie1} B. S. Xie, K. F. He, and Z. Q. Huang, Phys. Plasmas {\bf 6},
3808 (1999).

\bibitem{ros} M. Rosenberg and G. Kalman, Phys. Rev. E {\bf 56}, 7166 (1997).

\bibitem{kaw} P. K. Kaw and A. Sen, Phys. Plasmas {\bf 5}, 3552 (1998).

\bibitem{xie2} B. S. Xie, K. F. He, Z. Q. Huang, and M. Y. Yu, Phys. Plasmas 
{\bf 6}, 2997 (1999).

\bibitem{xie3} B. S. Xie and M. Y. Yu, Phys. Rev. E {\bf 62}, 8501 (2000).

\bibitem{sal} M. Salimullah and M. salahuddin, Phys. Plasmas {\bf 5}, 828
(1998).

\bibitem{cra1} N. F. Cramer, L. K. Yeung and S. V. Vladimirov, Phys. Plasmas 
{\bf 5}, 3126 (1998).

\bibitem{shu} P. K. Shukla and L. stenflo, Phys. Plasmas {\bf 7}, 2740
(2000).

\bibitem{xie4} B. S. Xie, Chin. Phys. Lett. {\bf 19}, 1463 (2002).

\bibitem{ich} S. Ichimaru, H. Iyetomi, and S. Tanaka, Phys. Rep. {\bf 149},
91 (1987).

\bibitem{chen} F. F. Chen, in {\it Plasma Diagnostic Techniques}, edited by
R. H. Huddlestone and S. L. Leonard (Academic, New York, 1965), Chap.~4.

\bibitem{tsy} V. N. Tsytovich, Phys. Usp. {\bf 40}, 53 (1997) [{\it Usp.
Fizicheskikh Nauk} {\bf 40}, 53 (1997)], and the references therein.
\end{references}
\end{document}